\documentclass[aps,prb,reprint,showpacs,longbibliography,superscriptaddress,nofootinbib]{revtex4-2}
\usepackage{bm}
\usepackage{graphicx}
\usepackage{epstopdf}
\usepackage{latexsym}
\usepackage{subfigure}
\usepackage{color}
\usepackage{hyperref}
\usepackage{amssymb}
\usepackage{amsmath}
\usepackage{amsbsy,bm}
\usepackage{fancyhdr}
\usepackage[T1]{fontenc}
\newcommand{\CsCoCl}{{$\rm Cs_2CoCl_4$\ }}
\begin{document}

\title{Antiferromagnetic resonance and two-magnon absorption in an XXZ-chain antiferromagnet {$\rm Cs_2CoCl_4$}}
\author{T.~A.~Soldatov}
\affiliation{P.~L.~Kapitza Institute for Physical Problems RAS, 119334 Moscow, Russia}

\author{A.~I.~Smirnov}
\affiliation{P.~L.~Kapitza Institute for Physical Problems RAS, 119334 Moscow, Russia}

%\author{A. V. Syromyatnikov}
%\affiliation{Petersburg Nuclear Physics Institute named by B.P. Konstantinov of National Research Center "Kurchatov Institute", Gatchina 188300, Russia}

\begin{abstract}

 Magnetic excitations of the exchange-dipole quasi 1D  XXZ antiferromagnet are studied in the ordered phase. We observe a transformation of
the electron spin resonance (ESR) spectrum when crossing the N\'{e}el temperature near 0.2 K. The single-mode ESR of a correlated XXZ chain transforms in the multi-mode spectrum
in the ordered phase. The multi-mode spectrum consists mainly of the intensive mode of a single correlated chain, which is surrounded and/or indented by numerous weak
satellites. The number of securely fixed modes is eight at ${\bf H}
\parallel b$ and twelve at ${\bf H} \parallel a$. Besides of the multi-mode resonance observed at the transverse polarization of the microwave and static magnetic fields, we
reveal a wide band of absorption by  ($k$,-$k$)- pairs of quasiparticles at the longitudinal polarization. This kind of absorption of microwaves occurs both in the ordered and
specific spin-liquid phases, revealing the presence of quasiparticles in the specific spin-liquid phase.

\end{abstract}

\date{\today}
 \maketitle

\section{Introduction}
\label{Introduction}

\CsCoCl~ is a quasi-1D magnetic  material  with an effectively anisotropic antiferromagnetic exchange. The spin structure is  built of Co$^{2+}$ ($S$=3/2)  magnetic ions. In a
low-temperature limit it  can be regarded  as a system of weakly interacting  chains of pseudospins $s$=1/2 \cite{BreunigPRL,Garst,Alvarez,Kenzelman}. The validity of the
pseudospin formalism is provided by the strong single-ion anisotropy. The scetch of the main exchange paths is given on Fig.1. The most strong exchange interaction $J$ acts
along $b$-direction of the distorted triangular lattice. The interchain exchange integral $J^\prime$ is much weaker according to the analysis of interatomic distances
\cite{Kenzelman} and experimental evaluation \cite{YoshizavaShiranePRB1983}. Complete detailed structure of the family of magnetic compounds Cs$_2$MX$_4$ (M=Co, Cu; X=Cl, Br) is
given clearly in, e.g., Refs. \cite{Kenzelman,BreunigPRB,PovarovPRR}.

\begin{figure}[t!]
\begin{center}
\vspace{0.1cm}
\includegraphics[width=0.42\textwidth]{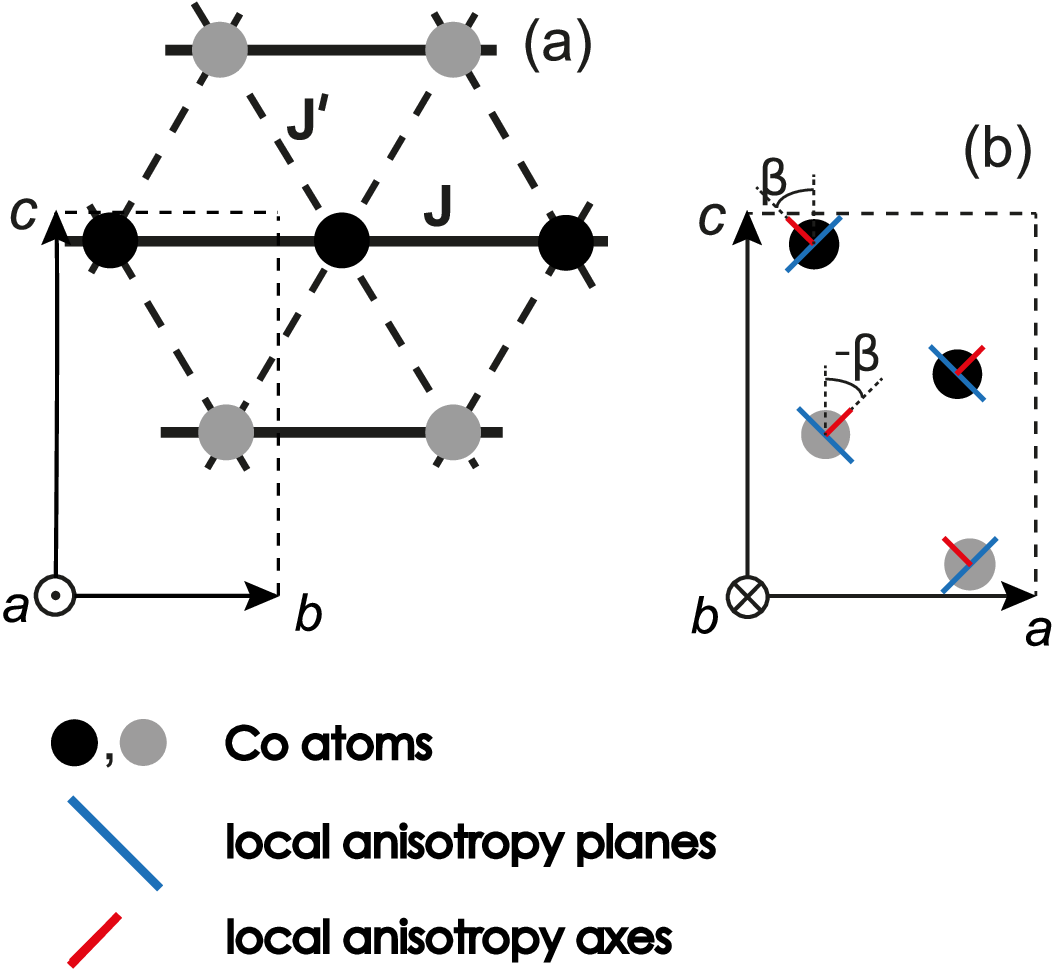}
\caption{Scheme of the main exchange paths in crystals of the family $\rm Cs_2MX_4$. Black and gray circles correspond to Co$^{2+}$-ions in the crystallographic positions $y=
\frac{1}{4}b$ and $y= \frac{3}{4}b$, respectively. Dashed lines and vectors of translations show the primitive cell.(a) Projection of the  structure on the $bc$-plane. (b)
Projection of the  structure on the $ac$-plane. Anisotropy axes and easy planes of Co$^{2+}$ ions are shown; $\beta\approx\pi/4$.\label{exchangenetwork}}
\end{center}
\end{figure}

\begin{figure}[h]
\begin{center}
\vspace{0.1cm}
\includegraphics[width=0.42\textwidth]{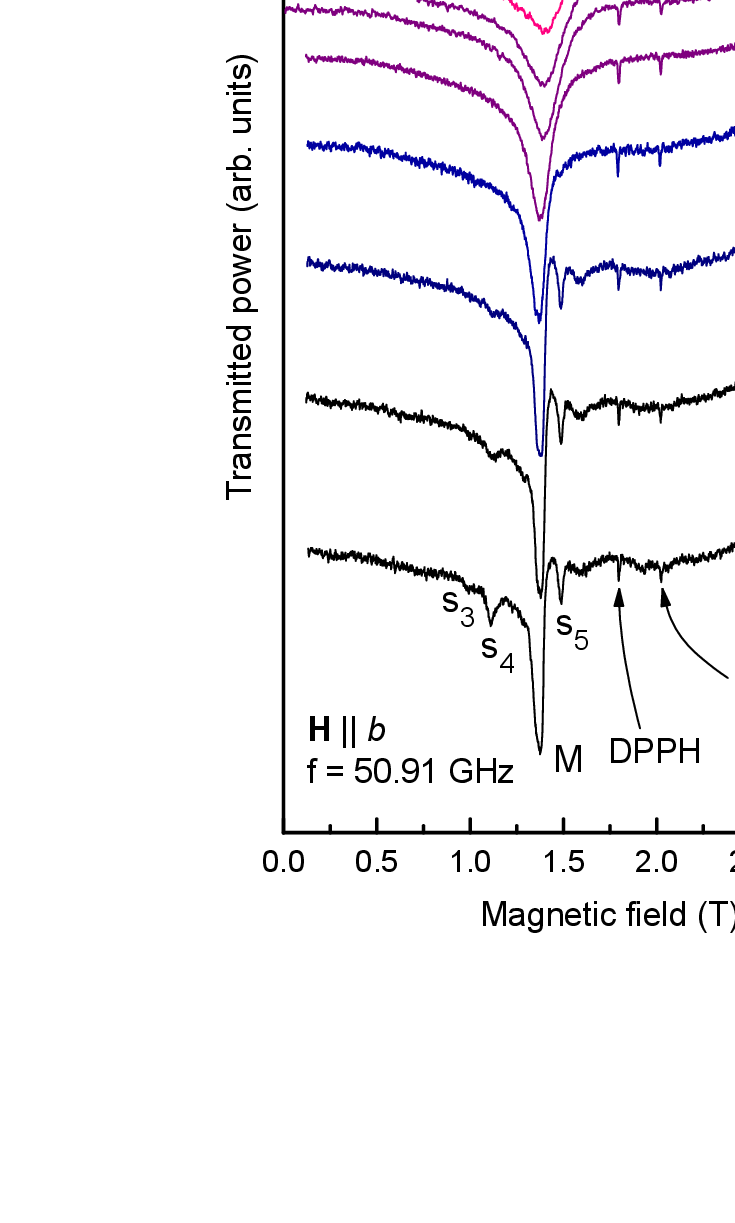}
\caption{\label{TdepHb} Temperature evolution of 50.91 GHz ESR lines of \CsCoCl at ${\bf H}\parallel b$. Letters indicate modes whose frequencies
are displayed on frequency-field diagrams in Fig.~\ref{FvsHb0p1K}. $H^*$ is the boundary of the absorption band, see text. The mixed polarization
of the microwave field ${\bf h}$ with respect to the static  field ${\bf H}$ takes place.}
\end{center}
\end{figure}

\begin{figure}[h]
\begin{center}
\vspace{0.1cm}
\includegraphics[width=0.42\textwidth]{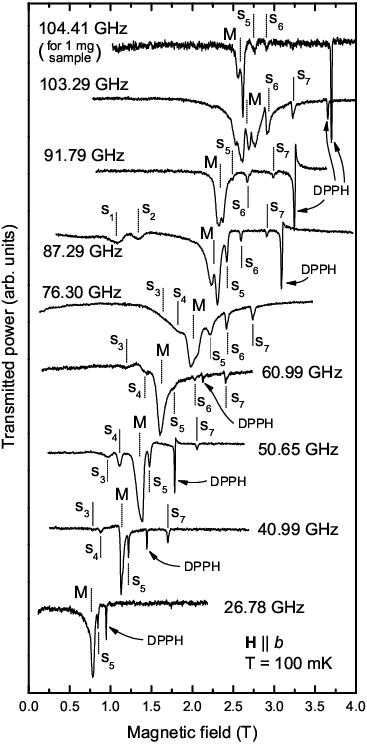}
\caption{\label{ESRlinesHb0p1K} ESR lines of \CsCoCl at $T=0.1$~K, ${\bf H} \parallel b$ and various frequencies. Letters indicate modes  whose frequencies are displayed on the
frequency-field diagram in Fig.~\ref{FvsHb0p1K}. The  polarization of the microwave field ${\bf h}$ at frequencies below 50 GHz is mainly perpendicular to the static field and
is of a mixed type at higher frequencies.  All curves except for the upper one are taken on a 2.2 mg sample. The upper curve presents the ESR line  for a smaller sample with a
weight of 1.0 mg and demonstrates the partial disappearence of the parasitic size effect present in the vicinity of the main ESR line M for a 2.2 mg sample, see text.}
\end{center}
\end{figure}
The  spin $S=1/2$ chain structure with an easy-plane or  easy-axis anisotropy is a case of the so-called XXZ chain.  The theoretical model of the antiferromagnetic $S=1/2$ XXZ
chain \cite{Garst,Alvarez,Kenzelman} reveals special ground states and quasiparticles. For the easy-plane anisotropy the quantum disordered (spin-liquid) ground state is
expected for a zero field. In a transverse (i.e. in-plane) field, the spin-flop antiferromagnetic phase with a strong reduction of the order parameter occurs. In a critical
field $H_c$ which is  still  below the saturation, the antiferromagnetic order should disappear and a spin liquid state should restore.  The saturation occurs only
asymptotically,  because of the lost of the axial symmetry in a transverse field. Besides, the noncommuting transverse magnetic field produces a strong entanglement in this
quantum magnet \cite{Alvarez}. The dynamics of XXZ chain is represented by a continuum like that of two-spinon excitations in Heisenberg or XY-antiferromagnet \cite{Alvarez,
Muller,kohnoPRL}. The continuum is changed for a gapped magnon mode in the magnetic field $H_c$. This evolution of the spectrum in a magnetic field  was found using DMRG
(density matrix renormalization group) modeling \cite{Alvarez,Garst} and confirmed qualitatively by inelastic neutron scattering (INS) \cite{Alvarez}.

To implement the transverse (noncommuting) orientation of the magnetic field one has to arrange  ${\bf H} \parallel b$ because of the alternation of the anisotropy axis between
the different chains, see Fig. \ref{exchangenetwork}.  For ${\bf H} \parallel b$  the field lies in the easy plane of each Co$^{2+}$  ion and thus is transversal to each local
anisotropy axis.

We can apply the model of $S=1/2$ XXZ  chain due to the strong single ion anisotropy of Co$^{2+}$ ($S=3/2$) ions. Because of the anisotropy, parametrized by the $D$-term in the
spin Hamiltonian, the upper $ S^z=\pm 3/2$ doublet is separated by a large gap of $2D$=14 K \cite{BreunigPRL} and at low temperature  these states  are not occupied. Because
only $S^z=\pm 1/2$ states are involved, the whole system may be described in terms of pseudospins $s=1/2$.

Thus, the initial $S=3/2$ Hamiltonian of a chain has a form

\begin{equation}
{\cal H} = \sum_i (J_{3/2} {\bf S}_i {\bf S}_{i+1} + D(S^z_i)^2+ g_{3/2} \mu_B   {\bf H S}_i ) \label{Ham}
\end{equation}
here $J_{3/2}$ is the exchange integral between $S=3/2$ spins, $D$ is the single ion anisotropy constant, $g_{3/2}$=1.9 is the primary nearly isotropic $g$-factor according to
experimental data of Ref. \cite{BreunigPRL}. For the transverse field ${\bf H} \parallel b$ and low temperatures  $T\ll D$ Hamiltonian (\ref{Ham}) may be replaced by the
following pseudospin representation (see, e.g., Ref. \cite{BreunigPRL}):

 \begin{equation}
{\cal H} = \sum_i (J_{1/2}(s^x_i s^x_{i+1} +  s^y_i s^y_{i+1} +\Delta s^z_i s^z_{i+1}) + g_{1/2}^x \mu_B H_x s^x_i ) \label{HamPseudo}
\end{equation}
here $s^\alpha _i$ are pseudospin $s=1/2$ component operators for site $i$ of the chain, $g_{1/2}^b$ is $g$-factor tensor component corresponding to $H_b$ magnetic field,
$x$-direction is along the crystallographic $b$-axis. The exchange integral gets a renormalization $J_{1/2}$=$4J_{3/2}$. For the limit $D \gg J_{3/2}$ we  have $\Delta$=0.25 and
$g_{1/2}^x$ =$2g_{3/2}$ . For the case of \CsCoCl when $J_{3/2}$=0.74 K  \cite{BreunigPRL, Kenzelman} and  $D$=7 K one has \cite{BreunigPRL} :
 $\Delta \simeq 0.25(1-39J_{3/2}/D$)=0.12  and $g_{1/2}^x \simeq
2g_{3/2}(1-\frac{3}{2} J_{3/2}/D$) =3.3. As follows from the analysis of thermodynamic experiments in a magnetic field \cite{BreunigPRL}, the  pseudospin $s=1/2$ representation
for \CsCoCl in the magnetic field is valid in a field range till 3.5 T.

The numerical simulations of the dynamics of an $S$=1/2 XXZ  chain in a magnetic field \cite{Garst, Alvarez} result in an excitation spectrum which is analogous to the spinon
continuum of a Heisenberg S=1/2 antiferromagnetic chain \cite{kohnoPRL}.  The difference between the XXZ and Heisenberg chains continua is a large field-induced gap between the
upper and lower boundaries of the XXZ chain continuum at $k$=0. At the same time the Heisenberg chain demonstrates close positions of these boundaries near the Larmor frequency. There is no essential difference for spectra calculated  at $\Delta$= 0.12 \cite{Garst} and $\Delta$=0.25 \cite{Alvarez}. The weak interchain interactions, e.g.,  $J^\prime$-exchange or
dipole-dipole interaction cause a 3D antiferromagnetic order at the N\`{e}el temperature $T_N$=0.22 K \cite{BreunigPRL,Kenzelman,BreunigPRB}, which is much lower than the
characteristic exchange temperature $T_{ex}=J_{1/2}$=2.9 K. At cooling through the temperature about $T_{ex}$ the in-chain correlations should appear. This is in the
agreement with the smoothed peak of the specific heat observed at $T$=1 K \cite{BreunigPRL}.

The specific heat of \CsCoCl  corresponds well to that of an ensemble of separate $S$=1/2 XXZ chains in the temperature interval from 0.3 to 1.5 K \cite{BreunigPRL}.
This is an additional reason to consider the spin system as noninteracting spin chains in this temperature range. At low temperature $T$ $\ll$ $T_{ex}$ the chains should be strongly intra-chain correlated and possess  properties of their ground state (g.s.). The theoretical investigation of the  temperature dependence of the spectra of $S$=1/2 XXZ chains \cite{Garst} shows the stable spectrum in the interval 0$<$ $T$ $<$ $T_{ex}/4$.  Thus we may conclude that at 0.35 K $<$ $T$ $\lesssim$ 0.7 K the chains  in \CsCoCl will be strongly intra-correlated and have a spectrum of excitations as in the g.s. Therefore we consider \CsCoCl in this interval as a special kind of a spin-liquid (further noted as a specific spin-liquid) with chains being close to the g.s. and without interchain coupling. As mentioned above, the  g.s. of a separate chain may be of the spin-liquid type or with a strongly reduced antiferromagnetic order. The specific spin-liquid is not a true spin-liquid but enables one to study the g.s. susceptibility of 1D spin system in 3D crystals. The spin-liquid spectra measured above the N\`{e}el temperature were noted earlier in study of spinons, e.g.,  in Cs$_2$CuCl$_4$ \cite{ColdeaPRL}.

\begin{figure}[htb]
\begin{center}
\vspace{0.1cm}
\includegraphics[width=0.5\textwidth]{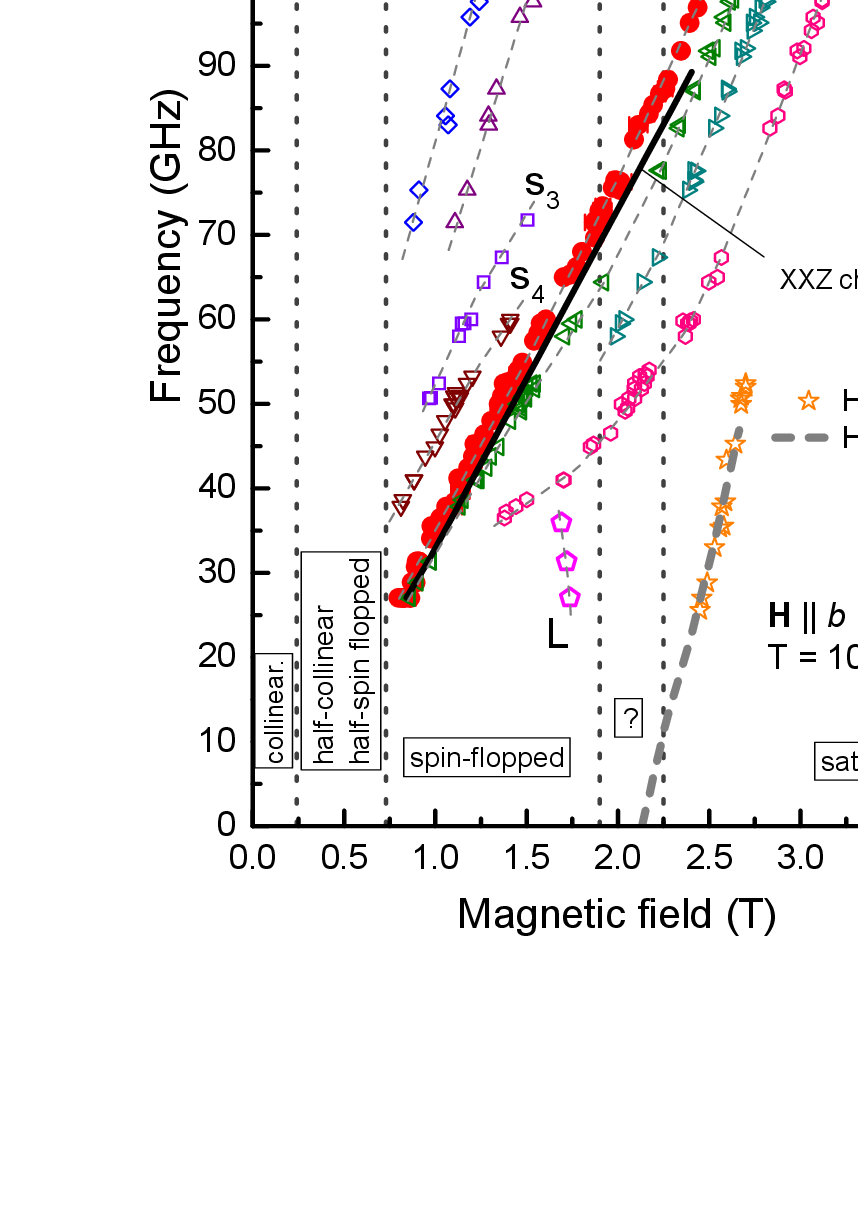}
\caption{\label{FvsHb0p1K} Frequency - field dependencies of antiferromagnetic resonance modes  of \CsCoCl at ${\bf H} \parallel b$ and $T=0.1$~K. The most intensive  ESR mode M
is presented by  closed symbols and weak resonance modes - by open symbols. Thin dashed lines are guide to the eyes. Vertical dashed lines denote the phase transitions according
to experimental results of Ref. \cite{BreunigPRB}. The boundary of absorption band $H^*$ observed at longitudional polarization of microwave field is presented by stars.Thick
grey dashed line presents the theoretical calculation of the field $H^*$, see text. Thick solid line is the theoretical frequency for XXZ chain with parameters of \CsCoCl. }
\end{center}
\end{figure}

\begin{figure*}[htb]
\begin{center}
\vspace{0.1cm}
\includegraphics[width=0.8\textwidth]{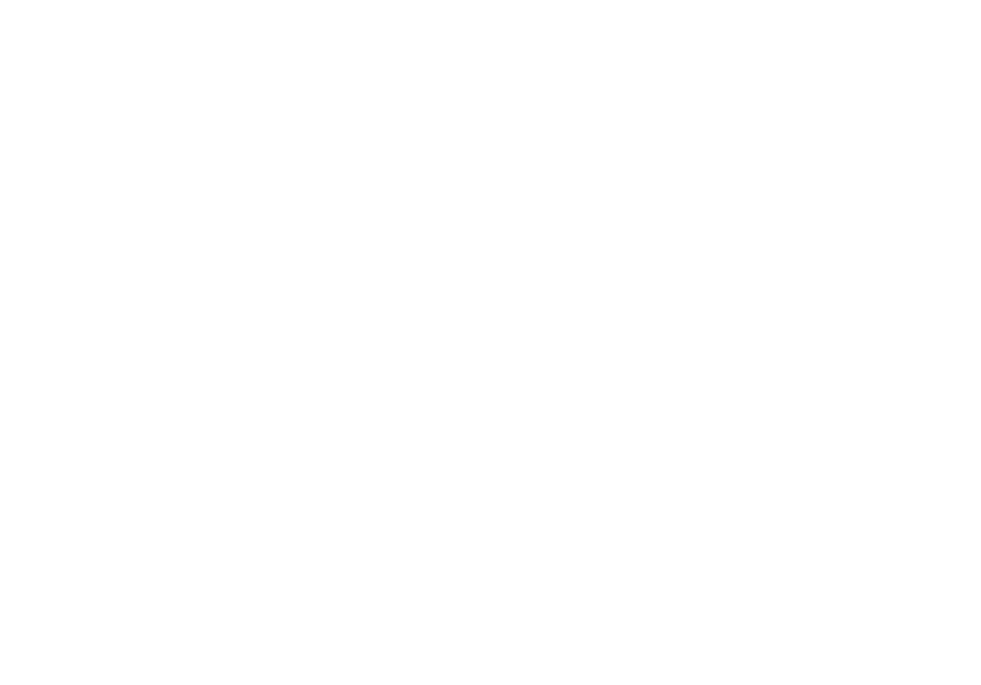}
\caption{\label{Tevolparandperp} Absorption curve temperature evolution in perpendicular (left) and parallel (right) polarizations of the microwave and static magnetic fields
 at ${\bf H} \parallel b$.
}
\end{center}
\end{figure*}

Basic magnetic properties and the phase diagram of ordered phases of  \CsCoCl are  studied in detail in Refs. \cite{BreunigPRL,BreunigPRB}, exchange and anisotropy parameters are
also derived from the experiment there. Neutron scattering  experiments \cite{YoshizavaShiranePRB1983} suggest the ratio of the interchain- to intrachain exchange 0.047. The
low-temperature elastic neutron scattering experiment Ref. \cite{Kenzelman} reveals the ordered eight-sublattice spin structure, which is almost collinear, with a tilting of
sublattices of about 15$^\circ$ to $b$-direction \footnote{Ref. [4] describes this structure as having 16  magnetic sublattices, this originates from Fig. 4 in Ref. [4], where a
sketch of a unit cell with 16 spins is given. This unit cell contains 2 primitive cells. The primitive cell may be presented as a rhomb containing only eight spins with the
numbers 3, 4, 6, 7, 9, 12, 13 and 14. The remaining eight spins may be obtained by moving the primitive cell via translations ${\bf (b+c)}$ and  ${\bf (-b+c)}$. Thus, 8 magnetic
sublattices are enough to describe this spin structure.}. This remarkable tilting is still not explained.

The phase diagram on the $H$-$T$ plane shows the paramagnetic phase, the  3D antiferromagnetic phase and  an additional phase of still unknown nature  in a narrow area between them \cite{BreunigPRB}. Presumably  this additional "phase 2"  may be a nematic phase \cite{BreunigPRB}. In the magnetic field ${\bf H} \parallel b$ it surrounds the N\`{e}el phase both from sides of higher fields and temperatures. Thus, in the magnetic field ${\bf H}$=1 T parallel $b$-axis  the transition to "phase 2" from the correlated XXZ-state occurs at $T_{c1}$=0.35 K and then to the N\`{e}el phase at $T_{c2}$=0.2 K.

\CsCoCl is a special member of Cs$_2$MX$_4$ family. All crystals of the family have the same crystal structure and exchange network. At the same time,  our target compound
\CsCoCl differs from the well known related spin chain system Cs$_2$CoBr$_4$ \cite{PovarovPRL2023,PovarovPRR,SoldatovSpinliq,SoldatovOrdered} by  a  more pronounced 1D
character, because the ratio of the interchain-  to the intrachain exchange for the last one is large ( of about 0.5  \cite{PovarovPRL2023}). It differs also from the Heisenberg
antiferromagnet  Cs$_2$CuCl$_4$ \cite{ColdeaPRL,ColdeaPRB,SmirnovPRL,SmirnovPRB2012} by essential anisotropy (they are XXZ and exchange-isotropic S=1/2 quasi 1D spin systems respectively).

The spectrum of spin excitations in the specific spin-liquid state of \CsCoCl was addressed previously by ESR (electron spin resonance) study \cite{ApplMagnRes2024} where the
pseudospin approach was confirmed experimentally via a direct measurement of the crucial $g$-factor 3.3 in a reasonable temperature range $T_{ex}<T<D$. At much lower
temperatures $T_N < T \ll T_{ex}$, experimental ESR frequencies in a wide magnetic field interval were found to be corresponding  to maxima of the calculated structure factor $S(k,\omega)$, which  describes the continuum of $S$=1/2 XXZ chain at $T$=0 \cite{Alvarez}. Because ESR absorption has a singularity at the same frequency as $S(0, \omega)$, this confirms spectroscopically a realization of strongly in-chain correlated but decoupled XXZ-chains. The evolution between the ESR of separate ions and collective resonance of $S$=1/2 XXZ-chain was observed as a smooth crossover between $T$=1.5 and $T$=0.6 K, i.e. within the paramagnetic phase.

In the present work we are looking for the change of the ESR  spectrum at the phase transition from a specific spin-liquid state to 3D  antiferromagnetic order near $T_N$. We see how
the spontaneous breaking of symmetry caused by a very weak interchain interaction changes the spectrum of spin oscillations. As a main result, we find a multimode spectrum with
an intensive mode at a frequency of the single chain mode surrounded by a set of weak satellites. Another unexpected observation is a finding of a band of two-quasi-particle
absorption present both in specific spin-liquid and ordered phases, showing the same nature of quasiparticles  in both phases.

\section{Experiment}
\label{exper}
    The ESR lines were  recorded as field dependencies of the microwave power transmitted through the resonator with the
sample at a fixed microwave frequency $\frac{\omega_{mw}}{2\pi}$. The sample was placed in a  maximum of the microwave magnetic field ${\bf h}$. In the vicinity of the ESR field
there is a diminishing of the transmission due to the absorption in the sample. The ratio of the signal $u_{tr}$ transmitted  through the resonator with the sample to the signal
$u_0$ passed through the empty resonator is \cite{Pool}:

\begin{equation}
\label{chiprimeprime}
 \frac{u_{tr}}{u_0} = \frac{1}{(1+2\pi \eta \chi^{\prime\prime} Q)^2}
\end{equation}

Here $\eta$ is the filling factor of the resonator and Q its quality factor, $\chi^{\prime\prime}$ is the imaginary part of the dynamic magnetic
susceptibility. For a small sample ($4\pi \eta \chi^{\prime\prime} Q \ll 1$) the change of transmitted power is proportional to
$\chi^{\prime\prime}$.

\begin{figure*}[htb]
\begin{center}
\vspace{0.1cm}
\includegraphics[width=0.8\textwidth]{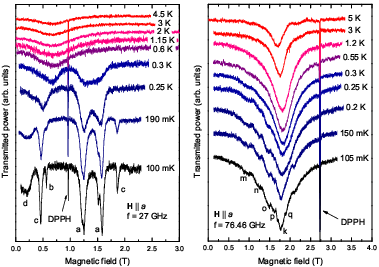}
\caption{\label{ESRvarTHa} Temperature evolution of ESR lines of \CsCoCl at 27.0  GHz and 76.46~GHz  for ${\bf H} \parallel a$. Letters indicate modes with frequencies presented
in Fig.~\ref{FvsHa01p1K}. }
\end{center}
\end{figure*}

\begin{figure}[htb]
\begin{center}
\vspace{0.1cm}
\includegraphics[width=0.42\textwidth]{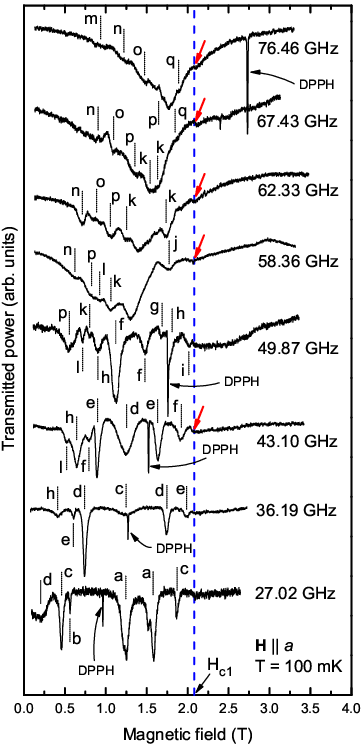}
\caption{\label{ESRvarfHa}  ESR lines of \CsCoCl at different frequencies for ${\bf H} \parallel a$ at T=0.1 K. Letters indicate modes which frequencies are presented in
Fig.~\ref{FvsHa01p1K}. Vertical dashed line marks the upper boundary field $H_{c1}$ of the canted antiferromagnetic phase according to Ref.\cite{BreunigPRB}. Tilted red arrows
show the kinks in the microwave signal in the field $H_{c1}$.}
\end{center}
\end{figure}

\begin{figure}[htb]
\begin{center}
\vspace{0.1cm}
\includegraphics[width=0.42\textwidth]{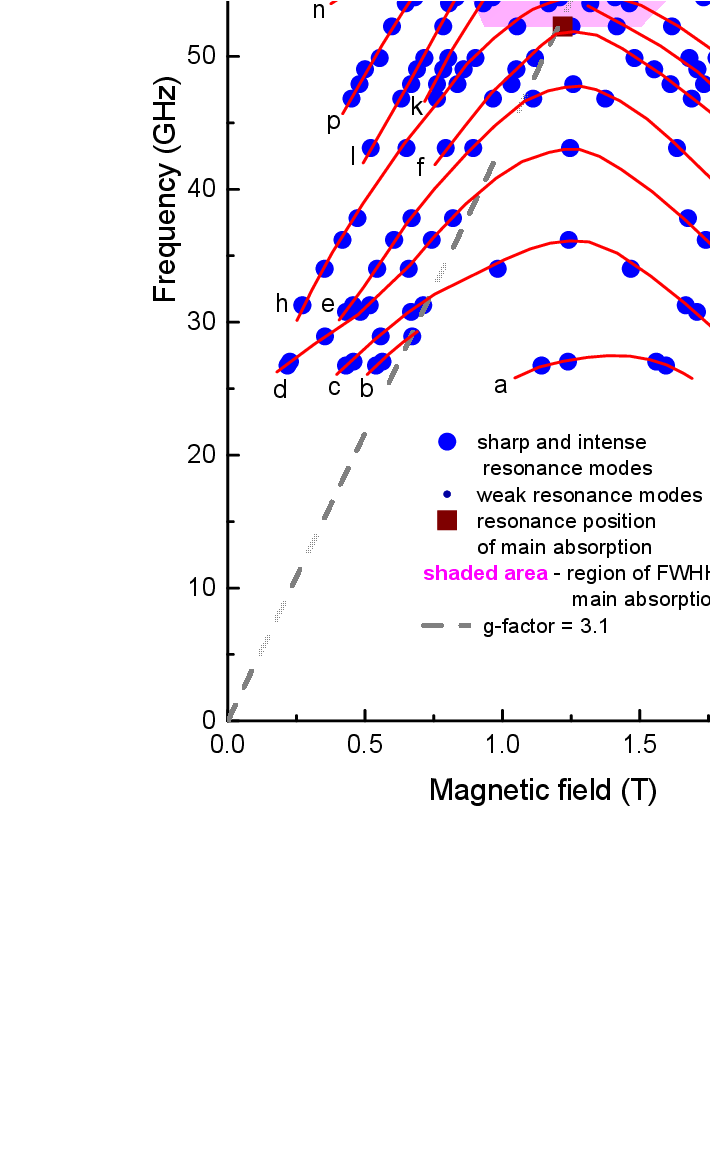}
\caption{\label{FvsHa01p1K} Frequency - field dependencies of antiferromagnetic resonance modes  of \CsCoCl for ${\bf H} \parallel a$ and $T=0.1$~K. The solid lines are guide to
the eyes. Tilted dashed line corresponds to paramagnetic ESR with $g=$3.1 observed at $T$=3 K.  Shaded area corresponds to the area within full width half height region of the
main ESR line. Vertical dashed line marks the field $H_{c1}$. }
\end{center}
\end{figure}

\begin{figure}[htb]
\begin{center}
\vspace{0.1cm}
\includegraphics[width=0.42\textwidth]{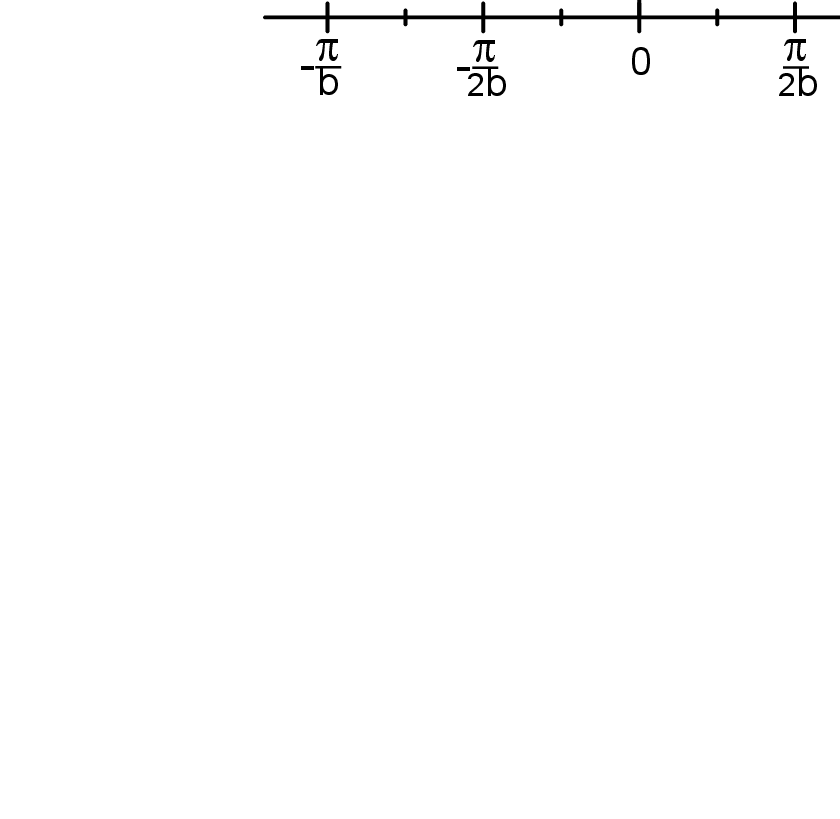}
\caption{\label{twoMagnonScheme}  Scheme of the two-magnon absorption: a photon of microwave radiation decays into two magnons with opposite
wavevectors. The field $H^*$ where the microwave frequency equals the double minimum frequency of the magnon spectrum is the boundary of the
absorption band. }
\end{center}
\end{figure}

We used a dilution microcryostat with sorption pumps combined with a microwave  resonator-type spectrometer to reach the low temperature of 0.1
K. Dilution cryostat and spectrometer constitute an insert into a dewar with 10 T cryomagnet \cite{Edelman}. The copper resonator of diameter 15
and height of 10 mm has about 40  discrete resonance frequencies in the range 25-120 GHz. For the comparison of the absorption in the
perpendicular and parallel polarizations we place the sample in different positions: at the bottom of the resonator for the perpendicular
polarization and at the wall for parallel one.  The axis of the resonator is parallel to the static field ${\bf H}$. A sapphire holder was also
used to change the position of the sample and control the polarization conditions of the experiment.

In  this research we deal with ESR modes of essentially different intensities. Therefore we have to use large samples for recording weak modes and small samples for intensive modes.
The intensive modes of large samples may cause a field-dependent detuning of the resonator  and parasitic size effects due to  field-dependent electrodynamic resonances within
the sample. These parasitic effects result in a strong deformation of the ESR lineshape, see, e.g. \cite{Smirnov2015CsCuCl}. The higher is the frequency the smaller should be
the sample to avoid the parasitic size effect. Thus, to record correctly the intensive lines we used smaller samples which do not distort the lineshape, but  do not allow detect
weak ESR lines due to the lack of sensitivity. For the control of the parasitic size effects we performed experiments on six samples with the masses ranging from  0.6 to 10 mg
and size between 1 and 3  mm.
   A small amount of 2,2-diphenyl-1-picrylhydrazyl (DPPH) was placed near the sample to get a $g=2.00$ marker.

\section{Experimental results}
\label{ExpResults}
\subsection{Experimental results for ${\bf H} \parallel b$}

 The evolution of the 50.9 GHz ESR line at cooling the sample from 4.5  to 0.1 K is shown in Fig. \ref{TdepHb}. These lines are recorded in an
experiment with a 2.5 mg sample on a sapphire holder in a position where the polarization of the microwave field is tilted. We observe a single line with $g$-factor 3.3 at 3$<T<$4 K, a superposition of two ESR lines at 1$<T<$2 K and then a single line in the field near 1.3 T at the temperature 0.3$<T<$0.8 K. This is the continuous crossover from
separate pseudospin ESR to a collective ESR mode of a correlated XXZ chain, studied in detail in Ref. \cite{ApplMagnRes2024}.
 At the further cooling through the N\`{e}el
temperature, which has the value 0.2 K in a field of 1 T \cite{BreunigPRB}, we observe a strong narrowing of the main line M and appearing of satellites S$_3$, S$_4$, S$_5$.
Besides of the separate ESR lines, we observe a wide band of absorption. A kink in the transmitted signal in a magnetic field 2.54 T marked as $H^*$ is a high-field boundary of
this absorption band. Below in subsection \ref{wideband} we will show that this band is most clearly observable if the microwave field ${\bf h}$ is parallel to the external
field ${\bf H}$. We do not observe essential changes in ESR response at cooling through the temperature $T_{c1}$=0.35 K, except for a continuous narrowing of the ESR line.

 For a set of frequencies in the range of 27 to 120 GHz the low-temperature ESR lines are presented in Fig. \ref{ESRlinesHb0p1K}.
 These lines are taken for a sample of mass 2.2 mg placed directly on the bottom of resonator, i.e. at a
perpendicular polarization of a microwave field. Experiments in the wide range reveal further satellites S$_1$, $S_2$, S$_6$, S$_7$. For the check whether the observed
satellites are singularities in the absorption of the target material and not due to parasitic size effects, we compare data taken on samples of different size. In Fig.
\ref{ESRlinesHb0p1K} this comparison is shown for 104 $\pm 1$ GHz ESR of samples with masses 1.0 and 2.2 mg, see two upper records. For a larger sample the main line M is sawtooth
distorted, thus a smaller sample should be used to study the main intensive ESR line, while for the weak satellite S$_7$ the large sample has a better signal-to-noise ratio. It
is worth to note that for this high frequency a weaker distortion of line M in a form of a splitting is still observable with a small sample. Here we determine the error in
the ESR field as the distance between two components of the bifurcated line. The results are summarized in the frequency-field diagram of Fig. \ref{FvsHb0p1K}. Here the
frequencies of the main line M and of seven satellites are given.  Note that most of the spectral weight belongs to the main line M. The Lorenzian fits of the observed ESR lines
transformed into $\chi^{\prime \prime}$ curves allow us to determine  the  values of the ratio of the sum of integral intensities of satellites to the integral intensity of line
M, given in Table 1.

\begin{table}[h]
\caption{ }
\begin{tabular}{|c|c|}
      \hline
f, GHz & $\sum I_{{\rm S}i}/I_{\rm M}$  \\ \hline

40.99 & 0.25 \\   \hline
43.77 & 0.19  \\   \hline
50.65 & 0.13  \\   \hline
60.99 & 0.07  \\   \hline
91.79 &  0.06  \\   \hline
\end{tabular}  \label{TableIntratioHb}
\end{table}

\subsection{Wide absorption band at ${\bf h} \parallel {\bf H}$}
\label{wideband}
 We observed a pronounced dependence of the absorption spectrum on the polarization of the microwave field ${\bf h}$. A comparison
of the 27 GHz ESR temperature evolution at perpendicular and parallel polarizations is presented in Fig. \ref{Tevolparandperp}. For a usual ESR polarization ${\bf h} \perp {\bf
H}$   there is a single ESR line or several well separated lines as seen on the left panel of Fig.\ref{Tevolparandperp}. For parallel polarization ${\bf h} \parallel {\bf H}$
the paramagnetic line  at $T>$ 2 K is not visible, while  a wide band of absorption arises below the temperature 2 K. This band has a well expressed high-field boundary at the
field $H^*$ visible in Fig. \ref{Tevolparandperp} (right panel) and Fig. \ref{TdepHb}. The band is observable very distinctly also in the ordered phase, below 0.2 K. In the
ordered phase the absorption band at frequencies 27-40 GHz is distorted in a chaotic manner in a left part of the band. For higher frequencies 45-53 GHz this chaotic distortion is not observable. On the background of this distortion the sharp resonances  M and L are visible, which were also
detected with the perpendicular polarization. One can see that these resonances have a strong polarization dependence: line M has a large intensity for ${\bf h} \perp {\bf H}$,
while line L appears mainly for ${\bf h} \parallel {\bf H}$.  The experimental values of the upper boundary $H^*$ revealed at different frequencies are shown in Fig. \ref{FvsHb0p1K} by stars.

\subsection{Experimental results for ${\bf H} \parallel a$}

The ESR in \CsCoCl exhibits a strong anisotropy. For  ${\bf H} \parallel a$ the temperature evolution of the ESR lines in the lower and upper parts of the frequency range is
presented in the left and right panels of Fig. \ref{ESRvarTHa} correspondingly. In the lower part of the range we see in the ordered phase several lines of resonance absorption
of the comparable intensity (lines d, c, a on the left panel of Fig. \ref{ESRvarTHa}.) At the same time, on upper frequencies (right panel) there is a single wide line weakly
distorted by sharp singularities. Some of these singularities do not have  a resonance shape. These weak features are marked on the 0.1 K record of 76.46 GHz ESR by letters m,
n, o, p, q. Examples of low-temperature ESR records on different frequencies are presented on Fig. \ref{ESRvarfHa}. Here we see that upon the frequency rise, the absorption
spectrum consisting  of several separate lines is transformed into a broad and intense line, distorted by weak resonances and weak irregular changes. The ratio of the sum of
integral intensities  in these weak features to the intensity of the wide strong 76.46 GHz line is about 10\%. Small steps are observed on  several ESR records in the magnetic
field $H_{c1}$=2.1 T, which is a high-field boundary of the canted antiferromagnetic phase determined in Ref. \cite{BreunigPRB}. The vertical dashed line is drawn through these
steps on Fig. \ref{ESRvarfHa}.

The frequency-field dependence of the described resonances and weak features of the absorption is displayed on Fig. \ref{FvsHa01p1K}. In the lower part of the frequency range
the spectrum has a ladder form with six well separated branches a, c, d, e, f, h.  At the frequencies above 50 GHz the spectrum transforms gradually in a central line containing
most of the spectral weight. The interval within the full width at the half height of this line is shadowed on Fig.\ref{FvsHa01p1K}. The frequency-field dependencies of the weak
features are given by black points connected by solid lines and marked by the same letters as on Figs. \ref{ESRvarTHa} and \ref{ESRvarfHa}.

\section{Discussion}
\label{disc}

\subsection{Magnetic field along {\it b}-axis}
\label{Hparallelb}

 For ${\bf H} \parallel b$ we note, that in a multi-mode spectrum  with 8 observed resonance modes only one mode is obviously
dominating, see Table I.  This line is accompanied by 7 satellites positioned above and below the frequency of the main line. The main line
corresponds qualitatively to  excitations  of the  XXZ  S=1/2 antiferromagnetic chain with the exchange and anisotropy parameters of Cs$_2$CoCl$_4$. In Fig. \ref{FvsHb0p1K} this correspondence is illustrated
by the black solid line  presenting the frequency of the lower maximum of $ k$=0 structure factor  of the  XXZ chain taken from DMRG calculations of Ref.
\cite{Alvarez} and the Supplemental Material of this Ref. For the rescaling of the magnetic field given in Ref. \cite{Alvarez} in energy units into a real magnetic field we use
the above value of $J_{1/2}$ and $g_{1/2}$=3.3. There is also an upper maximum of the structure factor $S(k,\omega)$, however, it is  much weaker and vanishing at $k$=0. In the field range 0.8-2.4 T there is a gap between the sharp lower and upper maxima of $S(k,\omega)$ at $k\rightarrow 0$. A continuum of frequencies arises far away from the center of Brillouin zone.  In the field range above 2.4 T (1.6 J in exchange units) calculation predicts a magnon-like  mode
in a whole Brillouin zone. Thus, only one ESR frequency is expected for a separate chain in the field range of the experiment. This mode has close frequencies to the main mode M, as shown in Fig. \ref{FvsHb0p1K}.

From a most general point of view, we may expect that the main features of  magnetic excitations in the ordered state of \CsCoCl should be mainly analogous to the excitations
 the XXZ chain and only weakly modified by the weak interchain interaction. The modification should result in a spectrum which should be,
on the one hand, close to the spectrum of excitations in a separate XXZ chain, and on the other hand, be a multimode spectrum of an eight-sublattice antiferromagnet. An
eight-sublattice antiferromagnet should have 8 modes of the antiferromagnetic resonance in the linear approximation  and may additionally  have composite excitations like bound
states of magnons, see, e.g., Ref. \cite{SoldatovOrdered}. Indeed, we see such a weak transformation of the single-mode spectrum to a multi-mode one, because the most of the
spectral weight remains near the ESR frequency of the XXZ chain and weak satellites provide the multicomponent structure of the spectrum.

The multi-mode antiferromagnetic resonance was observed earlier in the sister compound Cs$_2$CoBr$_4$ \cite{SoldatovOrdered}.  The Br-system in a low field range is a
collinear antiferromagnet in a longitudinal  field, while the Cl-system studied in present experiment is a flopped antiferromagnet.  Eight modes of absorption are observed in Cs$_2$CoBr$_4$, including intensive modes and weak satellites like here in $\rm Cs_2CoCl_4$. For the case of  Cs$_2$CoBr$_4$ a  microscopic theory in 4-sublattice (2D) approximation  was developed  \cite{SoldatovOrdered}, which describes satisfactory the number of
modes, their frequency-field dependencies, the intensive absorption by magnon modes, their weak satellites and a more weaker susceptibility of bound magnon states. Unfortunately, the results of this theory can not be immediately applied to ESR in \CsCoCl due to the difference of ground states and because of a vanishing interchain coupling in $\rm Cs_2CoCl_4$,
which is comparable in energy to the dipole-dipole interaction. This makes necessary to include the dipole-dipole interaction as a source of the interchain coupling and to
consider \CsCoCl as a rare example of a dipole-exchange antiferromagnet. This naturally makes the  theory more complicated.  Indeed, the characteristic dipole energy of a pair
of spins is $2(g_{1/2}\mu_B s)^2/r^3$=0.014 K, here $r\approx$ 6.2 \AA \   is a minimum distance between Co$^{2+}$ ions. At the same time the characteristic energy of the
interchain coupling expressed in terms of the interchain exchange is $J^\prime_{1/2}s^2$=0.035 K. Taking into account the long-range nature of the dipole-dipole interaction,
these estimations suggest its comparable contribution to the interchain coupling.
 According to Ref. \cite{SyromyatnikovPrivate}, namely  the dipole interaction in \CsCoCl results in the ground state
with  tilted sublattices, found experimentally in Ref. \cite{Kenzelman}.

Thus, from the multimode character of the ESR spectrum of \CsCoCl and the results of theory and experiment in Ref. \cite{SoldatovOrdered} we may suppose that the observed
intensive mode is of  magnon-type, which has weak satellites, some of them presumably are  bound states of magnons.

\subsection{Two-quasiparticle absorption}

Now we discuss the wide nonresonant band of absorption observed for a parallel polarization of the microwave field ${\bf h} \parallel {\bf H} $ both in the 1D-correlated (specific
spin-liquid) and ordered regimes at the static field along $b$-axis. The absorption band may be associated  with  the so-called two-magnon absorption (see, e.g,
\cite{ProzorovaSmirnov}) or parametric excitation of magnons (see, e.g., \cite{ProzorovaKveder}).  Both these types of absorption arise from the nonlinear coupling of the
microwave field with spin excitations. This coupling implies selection rules for a transformation of a photon into two magnetic quanta, i.e.
\begin{equation}
\label{selectrule}
 \omega_{mw}=\omega_k+\omega_{-k},
\end{equation}
(k-number of a microwave is practically negligible in comparison with $\pi/b$). The polarization for this coupling in antiferromagnets is typically parallel. The parametric
process was called for this reason "parallel pumping of magnons". The parametric excitation has naturally a threshold level of the microwave magnetic field. However, the
sub-threshold two-magnon absorption  also takes place and has a significant dynamic susceptibility in low dimensional antiferromagnets \cite{ProzorovaSmirnov}.

 A characteristic
sign of this kind of absorption is that it should occur in the field range where the relation (\ref{selectrule}) is fulfilled. Thus, for the boundary field we have
$\omega_{mw}=2(\omega_k)_{min}$, see the scheme of the boundary field calculation in Fig. \ref{twoMagnonScheme}. We  use the same spectrum \cite{Alvarez} of the XXZ chain  for the specific
spin-liquid phase and the ordered phase, as discussed above. This naturally results in the same edge field $H^*$ for the ordered and specific liquid phase.

The spectrum of the XXZ chain has a shape of a continuum in the field range below $H_c$=2.1 T. In higher fields the continuum shrinks to resonance magnon branch, see Fig. S11 in the
Supplemental Material  of Ref. \cite{Alvarez}. A shoulder in the absorbed power marking the upper boundary field of absorption $H^*$ is observed in the range where the continuum
is changed to a magnon branch. The $\omega$ vs $k$ dependence for this magnon branch is shown schematically in Fig. \ref{twoMagnonScheme} for three field values.
 From the calculated spectrum \cite{Alvarez} we take the minimum values of $\omega_k$ which appear at $k=\pi/b$. This minimum value depends
 on the magnetic field, starting from zero at $H$=$H_c$. The result of this calculation of $H^*$ is shown in Fig. \ref{FvsHb0p1K} and is in a good correspondence with observations presented by stars.
 There should be no left boundary of the band, because, according to Ref.\cite{Alvarez}, for all the fields below
$1.4J_{1/2}/(g_{1/2}\mu_B s)$=1.7 T there is a  continuum of absorption at $k=\pi/b$ on frequencies from zero to approximately 60 GHz. This  allows the two quasi-particle
absorption upto the zero field.

 We do not observe a threshold power for the absorption
band at the available microwave intensity. Probably, the absorption band is rather due to the sub-threshold absorption then to parametric excitation.

 Resuming this part we
again conclude that the spectrum of magnetic excitations in the ordered phase may be approximately (for about 90 \% of the spectral weight)
described as the spectrum of XXZ -chain.

   The appearance of the two-quasiparticle absorption both in the ordered and the specific spin-liquid phases shows that in the specific spin-liquid phase there are well
   defined quasiparticles, these should be spinon-like excitations of XXZ-chains in the fields below $H_c$ and  magnon-like excitations of the same chains in fields above $H_c$,
   as considered in Ref. \cite{Alvarez}.

\subsection{Magnetic field along {\it a}-axis}

 At frequencies below 60 GHz the frequency-field dependencies at ${\bf H}\parallel a$ form a ladder-type set of cupola, corresponding to modes a, c, d, e, f, h, k shown in
 Fig. \ref{FvsHa01p1K}. This differs essentially from the case of ${\bf H}\parallel b$ with monotonic frequency-field dependencies.
 The frequency-field diagrams at  ${\bf H} \perp b $  for both compounds \CsCoCl (this work) and Cs$_2$CoBr$_4$ \cite{SoldatovOrdered} are analogous  having this cupola-like shape.
 %{\it (Probably we have to discuss here whether this staircase is Zeeman ladder observed in C2CoBr4????)}
  On the frequencies above 60 GHz the spectrum of \CsCoCl may be again
 mainly represented by a single mode with a frequency close to the observed in the spin-liquid regime, see Figs. \ref{ESRvarfHa},
 \ref{FvsHa01p1K}.
The arising satellites are mainly within the width of the main mode and are of a weak intensity. This corresponds to the general point of view
 discussed in the subsection \ref{Hparallelb}- the high frequency part of the spectrum is close to that of XXZ chain.
We observe an essential change of ESR records at crossing the N\`{e}el temperature for low frequencies  $\omega/2 \pi \lesssim J_{1/2}/2\pi\hbar$= 60 GHz: several separate modes of comparable
intensities arise. The essential difference in the transformation of the spectrum at low and high frequencies is typical  for quasi 1D antiferromagnets.
At low frequencies in the N\`{e}el state the crystal may demonstrate excitations with energies comparable with interchain interaction energy.
%Here  the 3D magnon
%modes are observable at low frequencies because the 3D ordering occurs under the action of a weak interchain interaction,
At the same time for frequencies above $J$ the 1D
excitations dominate,  as observed, e.g. in a Haldane magnet CsNiCl$_3$ \cite{zaliznyak}
 and in an S=1/2 Heisenberg chain antiferromagnet Cs$_2$CuCl$_4$ \cite{ColdeaPRL,SmirnovPRB2012}.

\subsection{Comparison of INS and ESR data}

The spin excitations in the ordered phase of \CsCoCl were studied by the INS in Ref. \cite{Alvarez}. A continuum of excitations was observed which was interpreted as being corresponding to that calculated for a separate $S$=1/2 XXZ chain in the transversal field, i.e. for ${\bf H} \parallel b$. Nevertheless, to get this correspondence, the magnetic field was renormed because the orientation of the field in this experiment was ${\bf H} \parallel a$. The temperature of the INS experiment was deep in the  ordered phase $T$=0.07 K, this was also noted as a possible deviation of the observed and calculated spectra. For example, at  $H_a$=1.5 T and $k\rightarrow 2\pi$, which is equal to  $k\rightarrow 0$, the continous scattering with the energy change in the interval 0.1-0.5 meV  was found.
In contrast to the continuum detected by INS, the ESR spectra at ${\bf H} \parallel a $, T=0.1 K and in the same field $ H$=1.5 T present a set of close discrete resonances at frequencies below 60 GHz and a wide resonance absorption distorted by weak sharp resonances at higher frequencies.

Nevertheless, there is no contradiction between INS and ESR data. Indeed,
the INS data of Fig.2 in Ref. \cite{Alvarez} show that the INS  intensity at the transferred momenta k=0 and
2$\pi$ in a field $H_a$=1.5 T  is spread continuously in the energy range from 0.1 to 0.5 meV. The ESR experiment covers a part of this range between 0.1 and 0.33 meV (24-80 GHZ) and reveals there ten discrete ESR modes a part of which is superimposed with the main wide mode, see Fig.8. Presumably, the INS resolution is not enough to reveal close discrete modes separated by narrow intervals of about 6 GHz (0.025 meV). Therefore in the neutron experiment several close lines may look as a broad band of continuous absorption.The main intensive ESR line and extrapolation of the INS to $k=0$ practically coincide at frequencies above 60 GHz (0.25 meV). At these energies the INS and ESR spectra  coincide (with exception of weak satellites) and both investigations result in the conclusion on mainly analogous spectra of the ordered phase and  individual chains.

% From the INS experiment  in  Ref. \cite{Alvarez} it is concluded that the spectra of excitations are approximately the same for individual chains and for the 3D ordered  phase.
%It is also proposed that the spectra should be mainly analogous in both field orientations ${\bf H}  \parallel a $ and ${\bf H}  \parallel b $. The INS-data are taken deep  in
%the 3D ordered phase at $T \ll T_N$ and only for ${\bf H}\parallel a$, but they are interpreted by the 1D theory of a single chain in a transverse field ${\bf H} \parallel b$. The
%INS spectrum is   indeed mainly in agreement with that theory with a use of renormalized parameters. In ESR experiments we find a difference  between the single-mode spectrum of
%the specific spin-liquid phase and the multi-mode spectrum of the ordered phase. Despite the fact that the spectral weight of new branches is only of about 10 \% of total
%intensity, the new branches differ in frequency  and are very sharp, strongly differing from the lineshape of the specific spin-liquid resonance. Besides, we find a strong
%anisotropy of ESR frequencies in the low-frequency range.

\section{Summary and conclusion}
\label{conc}

Antiferromagnetic resonance in the ordered low-temperature phase of the pseudospin $s$=1/2 XXZ   dipole-exchange antiferromagnet  \CsCoCl is studied experimentally. We observe a
following  feature of the antiferromagnetic resonance of a quasi-1D antiferromagnet with a strong quantum reduction. At frequencies  above the exchange frequency $J_{1/2}/(2 \pi
\hbar)$ = 60 GHz the absorption spectra of the ordered phase are nearly the same as in the 1D correlated state (i.e. in the specific spin-liquid state). The most part of the spectral weight is in this quasi-1D
mode. Only weak satellites arise additionally  in accordance with the multi-sublattice ordering. For low-frequency excitations, the transformation of the absorption line at the transition through the N\`{e}el point is
more pronounced and the ESR line in the ordered phase is not similar to that of an XXZ chain, because the energy of interchain interaction is of the order of excitation energy.  Based on the theoretical analysis and experimental results for the sister compound Cs$_2$CoBr$_4$ we suppose that the observed ESR signals correspond to magnon modes of the 8-sublattice
structure and some of weak signals may be ascribed to bound states of magnons.

Besides, we have observed  a nonresonant band of two-quasiparticle absorption  both above and below the N\`{e}el temperature. The observation of this band in both phases
confirms the common 1D nature  of the  modes of spin oscillations in the specific spin-liquid and ordered phases. This confirms also the formation of quasiparticles in the
specific spin-liquid phase.

\begin{acknowledgments}

We are grateful to K.\ Yu.~Povarov for presenting the samples, V. N. Glazkov, S. S. Sosin,  L. E. Svistov and A.~V.~Syromyatnikov   for numerous stimulating discussions. We
acknowledge very sincerely the cooperation with V.~S.~Edelman  in use of the dilution microcryostat.  We acknowledge support by the Russian Science Foundation, Grant No.\
22-12-00259-$\Pi$ for ESR experiments in the high frequency range 60-130 GHz and polarization experiment  and State assignment of Kapitza Institute  for support of microwave
measurements in the low-frequency range and with small samples.

\end{acknowledgments}

\bibliography{litCCCv05Nov24}
\end{document}